\documentclass[doublecol]{epl2}

\title{Earthquake-like patterns of acoustic emission \newline in crumpled plastic sheets}
\shorttitle{Earthquake-like patterns of acoustic emission in
crumpled plastic sheets}

\author{R. S. Mendes\inst{1,2} \and L. C. Malacarne\inst{1} \and R. P. B. Santos\inst{1}
\and H. V. Ribeiro\inst{1} \and S. Picoli Jr\inst{1} }
\shortauthor{R. S. Mendes \etal}

\institute{
  \inst{1} Departamento de F\'\i sica, Universidade Estadual de
Maring\'a, Avenida Colombo 5790 \newline 87020-900,
Maring\'a-Paran\'a, Brazil\\
  \inst{2} National institute of
Science and Thechnology for Complex Systems, 22290-180 Rio de
Janeiro, RJ, Brazil\\
}


\abstract{We report remarkable similarities in the output signal of
two distinct out-of-equilibrium physical systems - earthquakes and
the intermittent acoustic noise emitted by crumpled plastic sheets -
Biaxially Oriented Polypropylene (BOPP) films. We show that both
signals share several statistical properties including the
distribution of energy, distribution of energy increments for
distinct time scales, distribution of return intervals and
correlations in the magnitude and sign of energy increments. This
analogy is consistent with the concept of universality in complex
systems and could provide some insight on the mechanisms behind the
complex behavior of earthquakes.}

\begin{document}

\maketitle

Understanding the underlying mechanisms that govern the complex
spatio-temporal behavior of earthquakes is a stimulating
challenge\cite{turcotte95,kagan99}. Concepts and methods from
statistical physics has been largely applied to study earthquakes,
contributing to identify several patterns in seismic
activity\cite{bak02,mega03,abe04,corral05,saichev06,caruso07,lennartz08,balankin09,abe09}.
This approach also has been contributing to identify universal
behavior in earthquakes - similarities between seismic records and
the output signal of systems in different research areas.

\begin{figure}
 \centering
 \DeclareGraphicsRule{ps}{eps}{}{*}
\includegraphics[scale=.52]{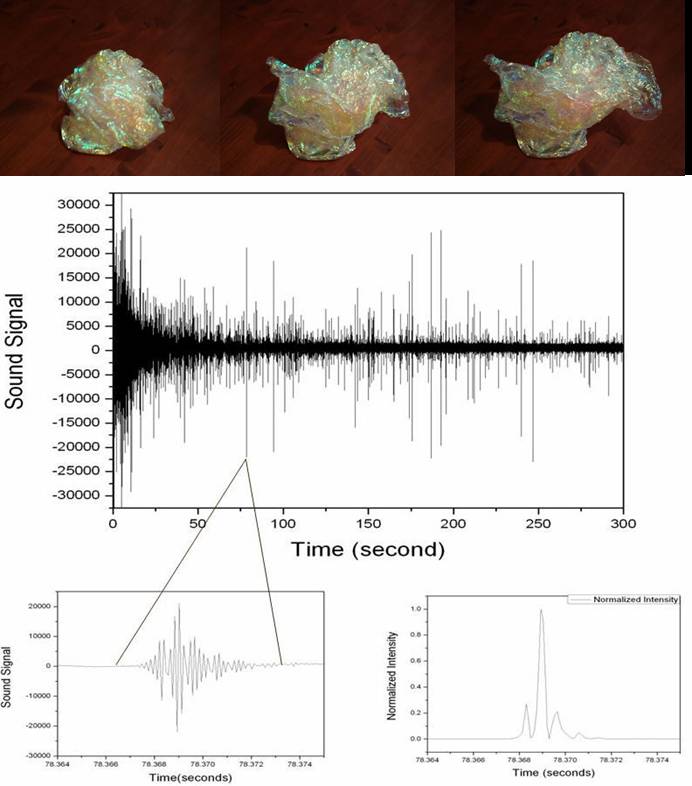}
 \caption{Obtaining the energy intensity from the
recorded acoustic noise. Top: A sample of crumpled plastic sheet in
relaxation. Middle: The corresponding acoustic noise recorded (in
arbitrary units). Bottom-left: Detail of a single event; and
Bottom-right: the corresponding intensity of a single event,
$I=cp^2$, where $c$ is a constant and $p$ is the sound pressure. The
variable $S$, given by the area under the corresponding set of
peaks, is related to the energy dissipated in the event. Because of
the short time distance between events in the beginning of the
relaxing process, it was not possible to discriminate individual
events. Then, for each experimental dataset, we cut out the first 10
seconds in the recording.} \label{fig1}
\end{figure}

For example, it has been reported that $\gamma$-ray events emitted
by neutron stars and earthquakes share several distinctive
statistical properties indicating tectonic activity on neutron stars
- 'starquakes' - analogous to earthquakes on Earth\cite{cheng95}.
Another example is a reported analogy between earthquakes and the
Internet. Specifically, it has been found that two known empirical
power laws for earthquakes - the Omori law and the Gutenberg-Richter
law - hold also for the Internet (ping experiment). In this context,
sudden drastic changes of the Internet time series are referred as
'internetquakes'\cite{abe03,abe04s}. Earthquake patterns also can be
observed in financial markets. It has been reported, for instance,
that stock price fluctuations follow a power law distribution with
exponent $\approx 3$\cite{gabaix03}. This behavior is quantitatively
similar to those found in earthquakes (see refs.
\cite{abe03,caruso07}), indicating an analogy between natural and
financial earthquakes. For other examples of universal behavior in
complex systems, see refs. \cite{ex1,ex2,ex3,ex4,ex5,ex6}.

Here, we compare earthquakes with the output signal of an
out-of-equilibrium physical system - the intermittent acoustic noise
emitted by crumpled plastic sheets. Some out-of-equilibrium physical
systems emit crackling noises as a response to external conditions
through events spanning a broad range of
sizes\cite{r1,r2,r3,r4,r5,r6,r7,r8}. In particular, the sharp and
intermittent noises emitted by some kinds of crumpled papers and
similar materials - including plastic sheets - qualitatively
remember earthquakes which arise when two tectonic plates rub each
other. Starting from this qualitative picture, we search for a
quantitative support for this analogy. Specifically, in a series of
experiments we measure the acoustic noise emitted by a crumpled
plastic sheet - a Biaxially Oriented Polypropylene (BOPP) film - in
relaxation and compare these records with real data on the magnitude
of earthquakes. We find that both processes exhibit several similar
statistical properties.

To quantitatively test this analogy, we consider real data on
earthquakes obtained from the Northern California catalog for the
period 1966-2006\cite{r9}. This seismic database contains $\sim
435,000$ records from one of the most active and studied geological
faults on the Earth - the San Andreas Fault. For each event, we
calculate a measure of the energy dissipated, $E=\exp(M)$, where $M$
is the reported magnitude of the earthquake.

We also perform a series of experiments in order to obtain measures
of the energy dissipated by crumpled plastic sheets in relaxation
process. First we crumpled a given sample of plastic sheet (of size
$1.0m$ x $1.5 m$) into a compact ball. This procedure is similar to
the common experience of crumpling an unwanted sheet of paper prior
to disposing of it\cite{r3}. In such conditions, the plastic sheet
emit sound in discrete pulses of a variety of intensities for a
relatively large time after released to relax (about 10 minutes).

We record the sound emitted with a condenser microphone (Shure
Microflex $MX202W/N$) positioned at 1 meter from the sample for five
minutes and digitalized at frequency 8000 Hz. For the analysis, the
noise was reduced by applying a cutoff filter for low frequencies. A
single event is identified by a set of peaks with intensity larger
than a threshold value. The start time of the single sound is taken
as the time corresponding to the first sound with intensity larger
than the threshold, whose value was chosen above the noise
intensity. The end time corresponding to the time when the intensity
becomes lower than the threshold for a time bigger than a given
$t_{c}\approx 1-2 ms$ (the characteristic length of single event).
Figure 1 shows a typical example of the acoustic emission recorded
and the corresponding energy intensity $S$ obtained for a specific
event.

We first determine the probability distribution of the energy
intensity for earthquakes and crumpled plastic sheets. Figure 2a
shows that both distributions are consistent with a power law decay,
\begin{equation}\label{eq1}
    P(x)\sim x^{-\alpha},
\end{equation}
with $\alpha\simeq 3$. This result, already known for earthquakes,
suggests that both systems are self-organized into a scale-free
state - there is not a typical scale for the dissipated energy.
Observe that the power law exponent is quantitatively similar for
earthquakes and crumpled plastic sheets.

\begin{figure*}[!ht]
 \centering
 \DeclareGraphicsRule{ps}{eps}{}{*}
\includegraphics[scale=.3]{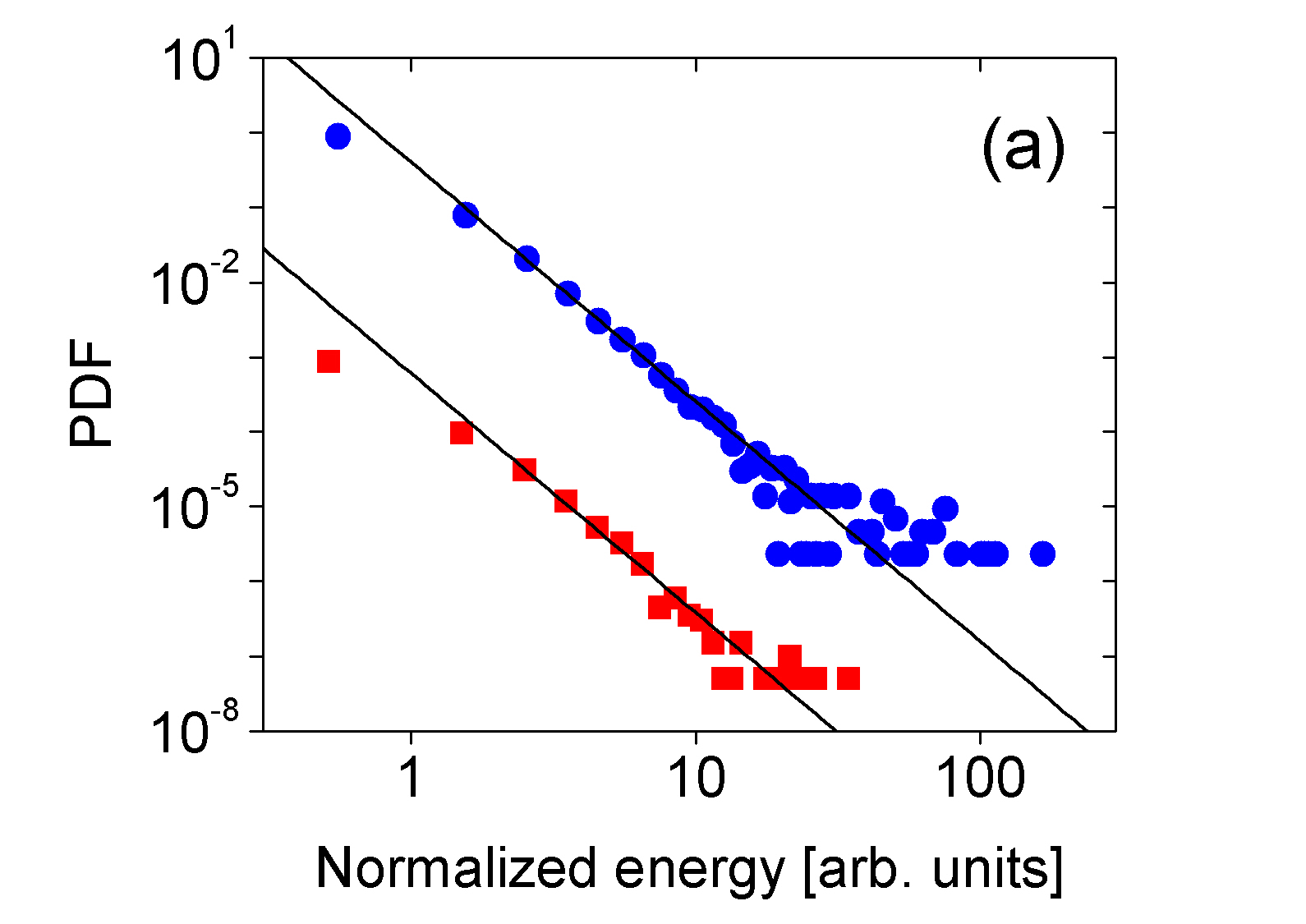}
\includegraphics[scale=.3]{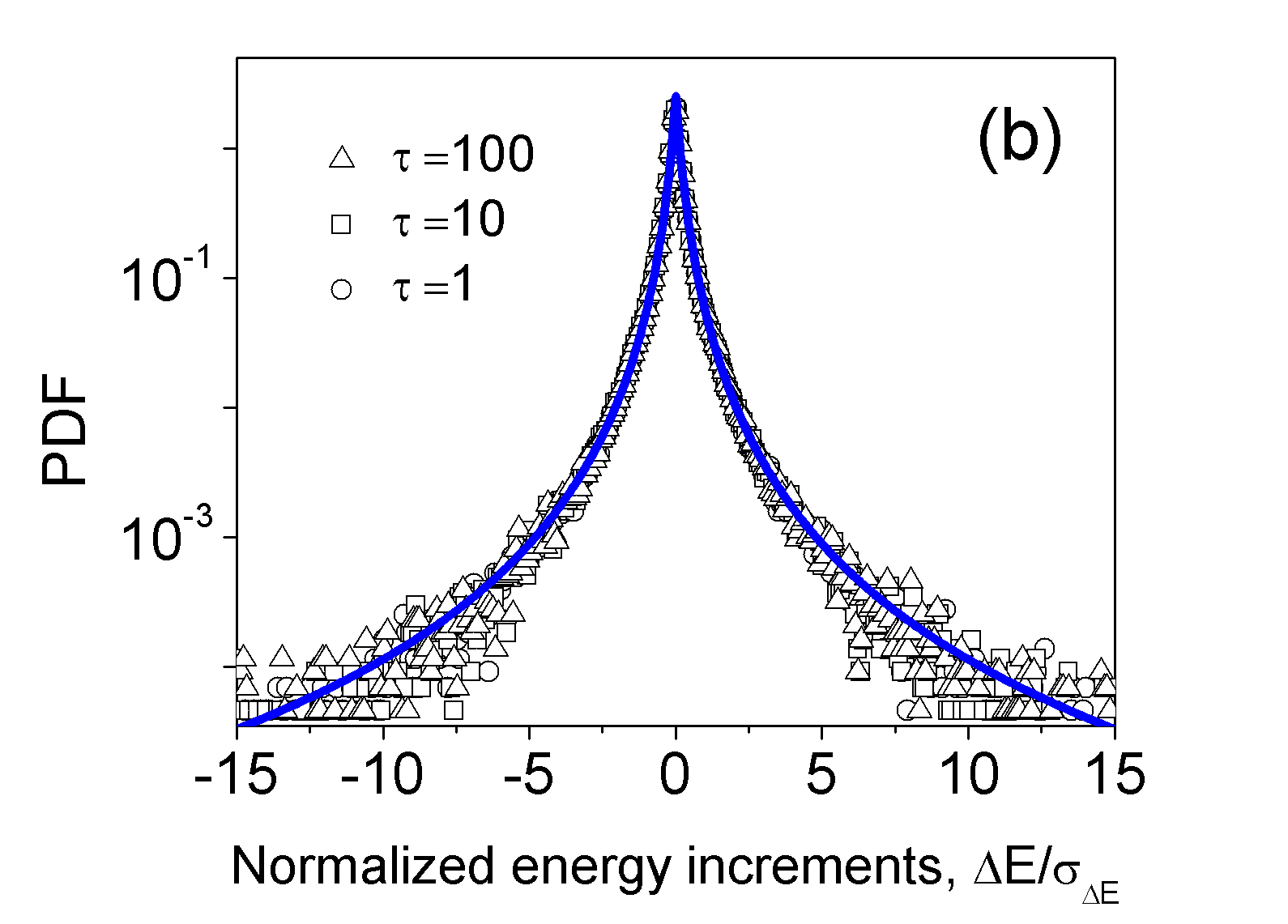}
\includegraphics[scale=.3]{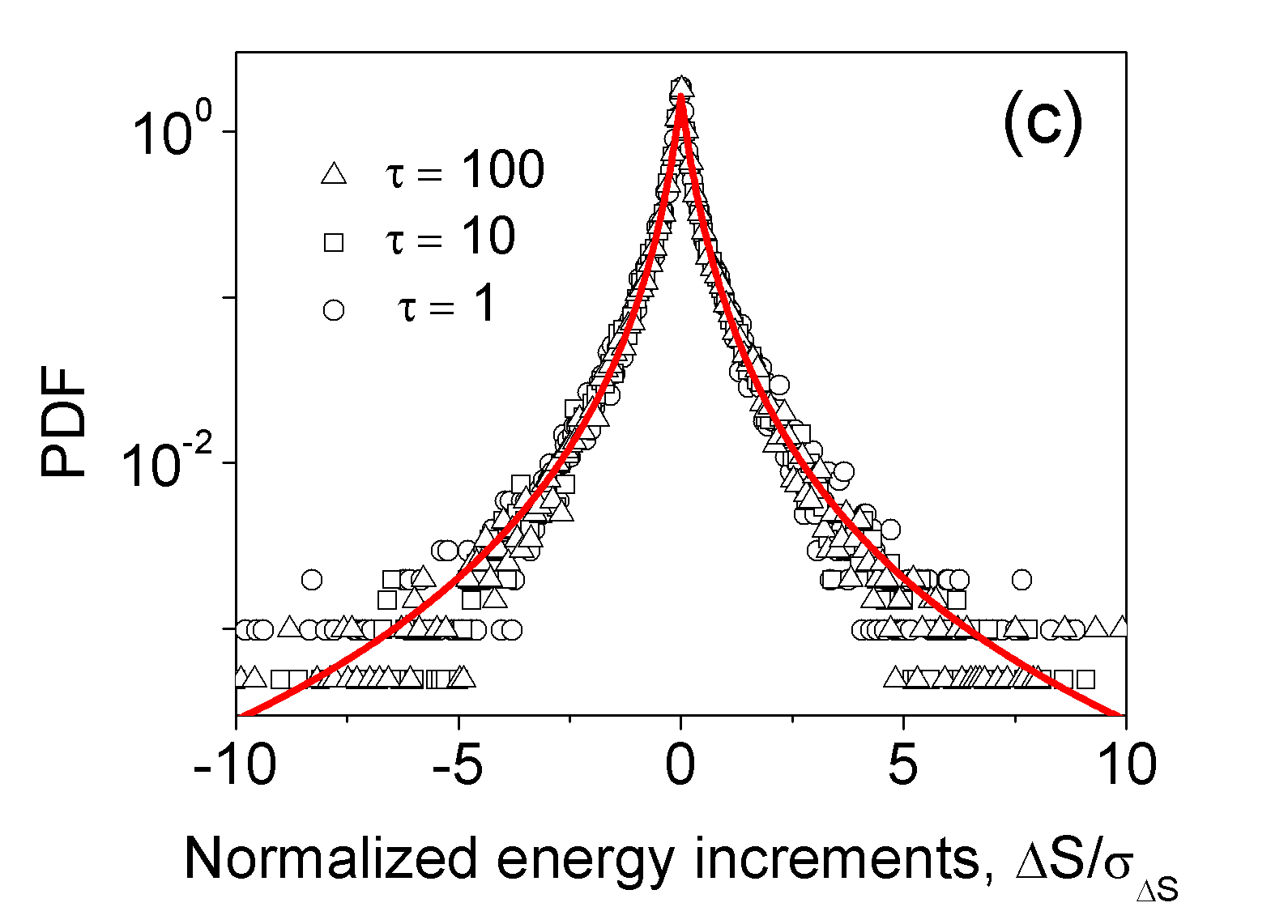}
\includegraphics[scale=.3]{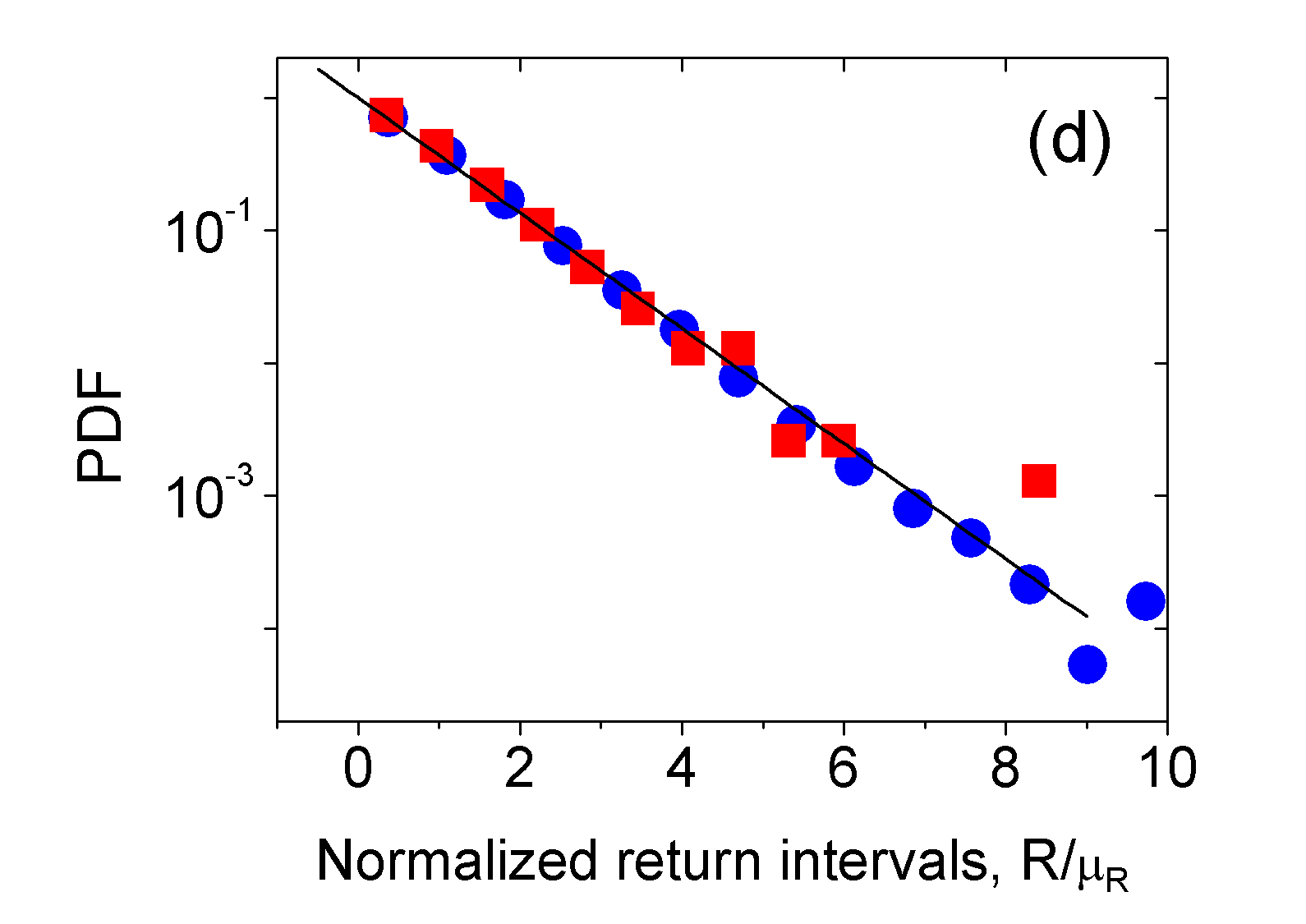}
 \caption{Analysis of the probability density function (PDF) of data.
 {\bf (a)} Probability density, $P(E/\sigma_{E})$ and $P(S/\sigma_{S})$, of normalized
 energy for earthquakes (circles) and crumpled films (squares). $\sigma$ is the standard
 deviation calculated over all records in a given series. The curves are shown vertically shifted
for clarity. The solid lines are power laws given by eq.
(\ref{eq1}), with exponent $\alpha=3.2$. {\bf (b)} Probability
density of normalized increments, $P(\Delta E/\sigma_{\Delta E})$,
for time scales $\tau=1, 10, 100$. The solid line corresponds to
equation (\ref{eq2}) with $\alpha=3.2$. {\bf (c)} Probability
density $P(\Delta S/\sigma_{\Delta S})$, for time scales $\tau=1,
10, 100$. As well as in b, the solid line is given by equation
(\ref{eq2}) with $\alpha=3.2$. {\bf (d)} Probability density of
normalized return intervals, $P(R/\mu_{R})$, for earthquakes
(circles) and crumpled films (squares). In both cases, $R$ is
calculated within subseries of size 1016. $\mu_{R}$ is the average
of $R$ in a given subseries. The threshold is $R_{c}=2$. The solid
line is an exponential distribution $P(r)=\exp[-r]$.} \label{fig2}
\end{figure*}

To find information on the dynamics of energy dissipation, we define
energy increments as $\Delta E=E(i+\tau)-E(i)$ and $\Delta
S=S(i+\tau)-S(i)$, where $E(i)$ and $S(i)$ are proportional to the
energy of i-th event. The distributions of energy increments, for
different values of the time scale $\tau$, are shown in Figures 2b
(earthquakes) and 2c (crumpled plastic sheets). All curves are
symmetrical, very peaked and have wings larger than expected for a
normal process. In both cases, data for distinct time scales
collapse onto a single curve indicating that the distribution of
energy increments exhibits a common functional form for all time
scales in the range considered. We also shuffled the original series
and then calculated the distribution of $\Delta E$ and $\Delta S$
again, but no significant changes were observed. This result may
indicate no correlations or weak correlations in the time
organization of energy intensities.

Assuming a given variable $x$ following a power law distribution
with exponent $\alpha$ (see eq. (\ref{eq1})) and no correlation
between two events (a first approximation), the probability
distribution for the increments $\Delta x=x(i+\tau)-x(i)$ is given
by $P(\Delta x)=K
    \int_{0}^{\infty}dx\int_{0}^{\infty}dx'(xx')^{\alpha}\delta(x'-x-\epsilon)=
    K \int_{\tau}^{\infty}dx[x(x+|\epsilon|)]^{-\alpha}$, where $K$ is a normalization
constant and $\epsilon$ is a small positive value to avoid
divergence in $x=0$. The integration leads, for real and positive
$\alpha$, to the normalized probability density function (PDF)
\begin{equation}\label{eq2}
    P(\Delta x)=\frac{(\alpha-1)^2}{\epsilon(2\alpha-1)}F_{1}^{2}\left(\alpha,2\alpha-1,2\alpha,-\frac{|\Delta
    x|}{\epsilon}\right),
\end{equation}
where $F_{1}^{2}$ is the confluent hypergeometric
function\cite{caruso07}. This PDF is shown in Figures 2b and 2c in
comparison with real data. Notice that both curves are given by eq.
(\ref{eq2}) with the same parameters. The good adjustment to the
data indicates that the distribution of energy increments exhibits a
common shape for both systems.

The non-Gaussian behavior of the distributions of energy increments,
shown in Figures 2b and 2c, also can be characterized by q-Gaussian
distributions - typical in Tsallis
statistics\cite{tsallis88,mendes98,picoli09s}. In fact, it has been
reported that eq. (\ref{eq2}) can be very well reproduced by means
of q-Gaussians, whose values of $q$ are related with the parameter
$\alpha$\cite{caruso07}. In the range considered, a q-Gaussian
distribution, with $q\simeq 1.75$, practically coincides with the
curves shown in Figures 2b and 2c (solid lines). For $q>1$, the
tails of a q-Gaussian decreases as a power law with exponent
$\beta=2/(q-1)$. This result indicates that the tails of the
distribution of energy increments follows a power law behavior,
\begin{equation}\label{eq2a}
    P(\Delta x)\sim \Delta x^{-\beta},
\end{equation}
with $\beta \simeq 2.7$ for both systems.

Another way to characterize the dynamics of the output signal of a
given system is to analyze the return interval series. The return
intervals $R$ are defined as the interval between events that exceed
a certain threshold $R_c$. We obtain $R$ from the normalized energy
series - with elements $E/\sigma_{E}$ and $S/\sigma_{S}$, where
$\sigma$ is the standard deviation. Figure 2d shows the distribution
$P(r)$ for earthquakes and crumpled plastic sheets, where
$r=R/\mu_{R}$ and $\mu_{R}$ is the average of $R$ in a given
subseries of size 1016 (the typical size of a given experimental
record for a crumpled sheet). For comparison, we also show the
exponential distribution $P(r)=\exp[-r]$. Observe that the
distribution of return intervals for earthquakes and crumpled films
share a common shape. We perform a parallel analysis for several
values of the threshold $R_c$ but no significative changes were
observed. The exponential behavior found in the distribution of
return intervals suggests no correlation or weak correlations in the
energy series of both systems.

\begin{figure*}[!ht]
\label{dfa}
 \centering
 \DeclareGraphicsRule{ps}{eps}{}{*}
\includegraphics[scale=.22]{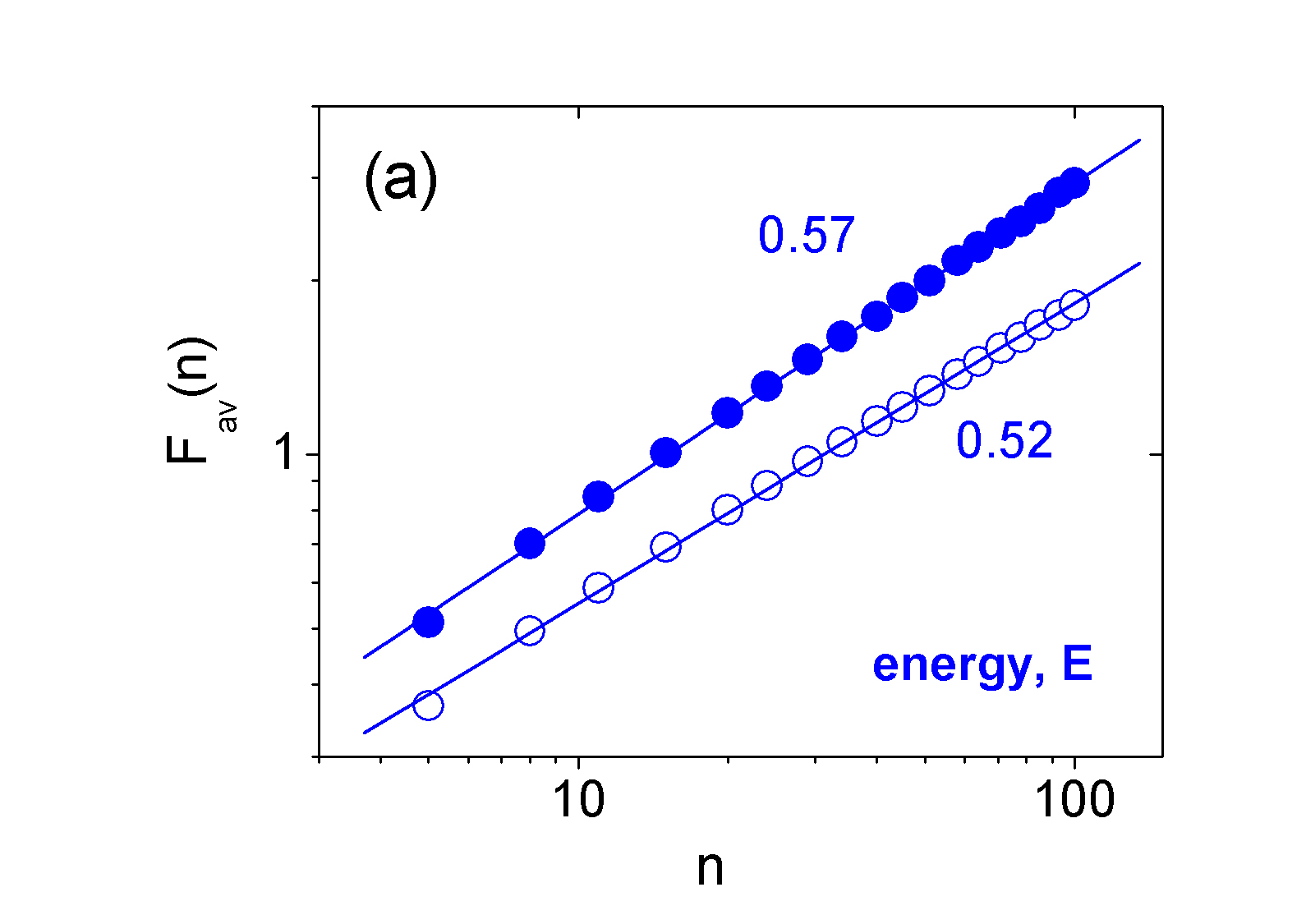}
\includegraphics[scale=.22]{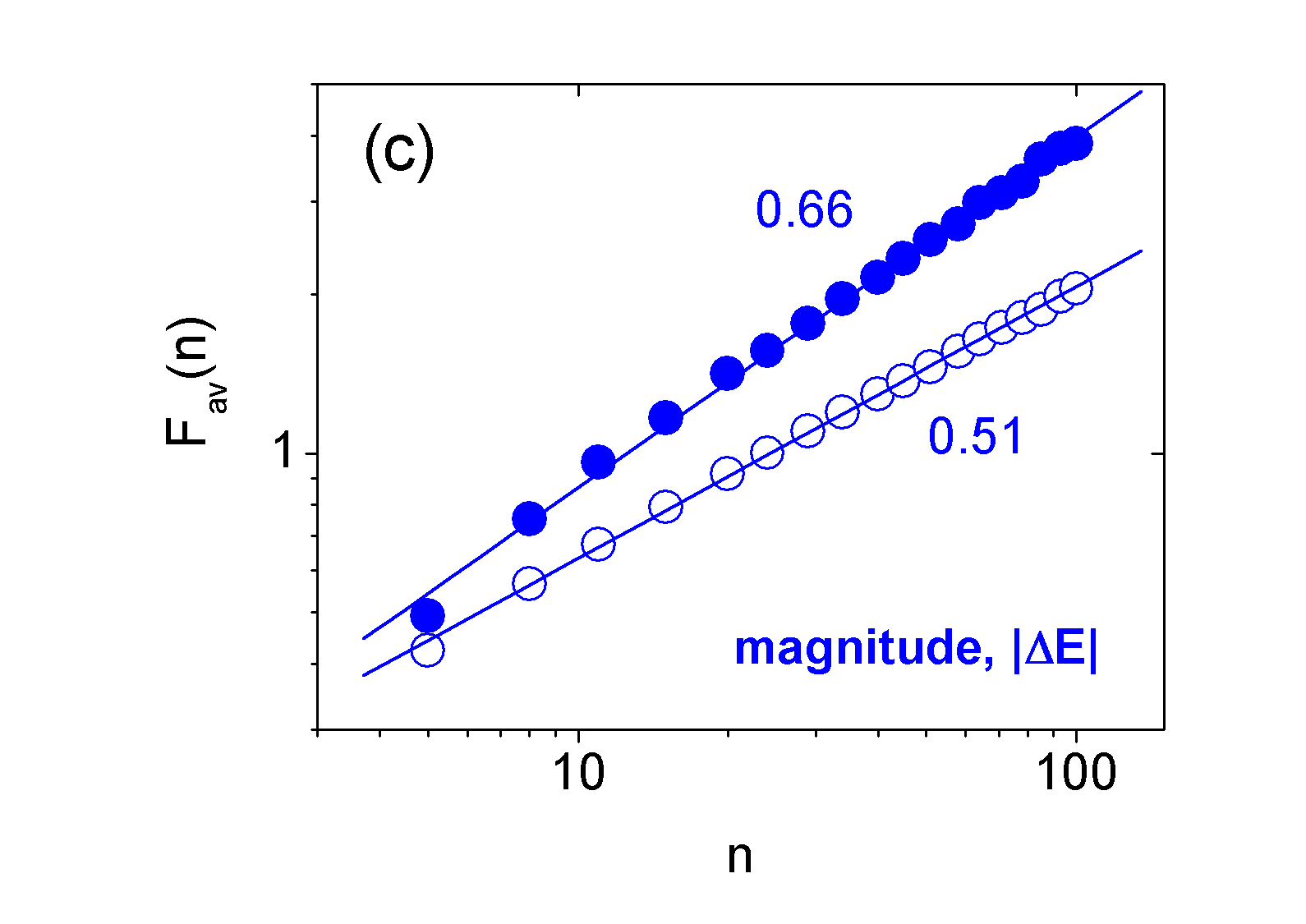}
\includegraphics[scale=.22]{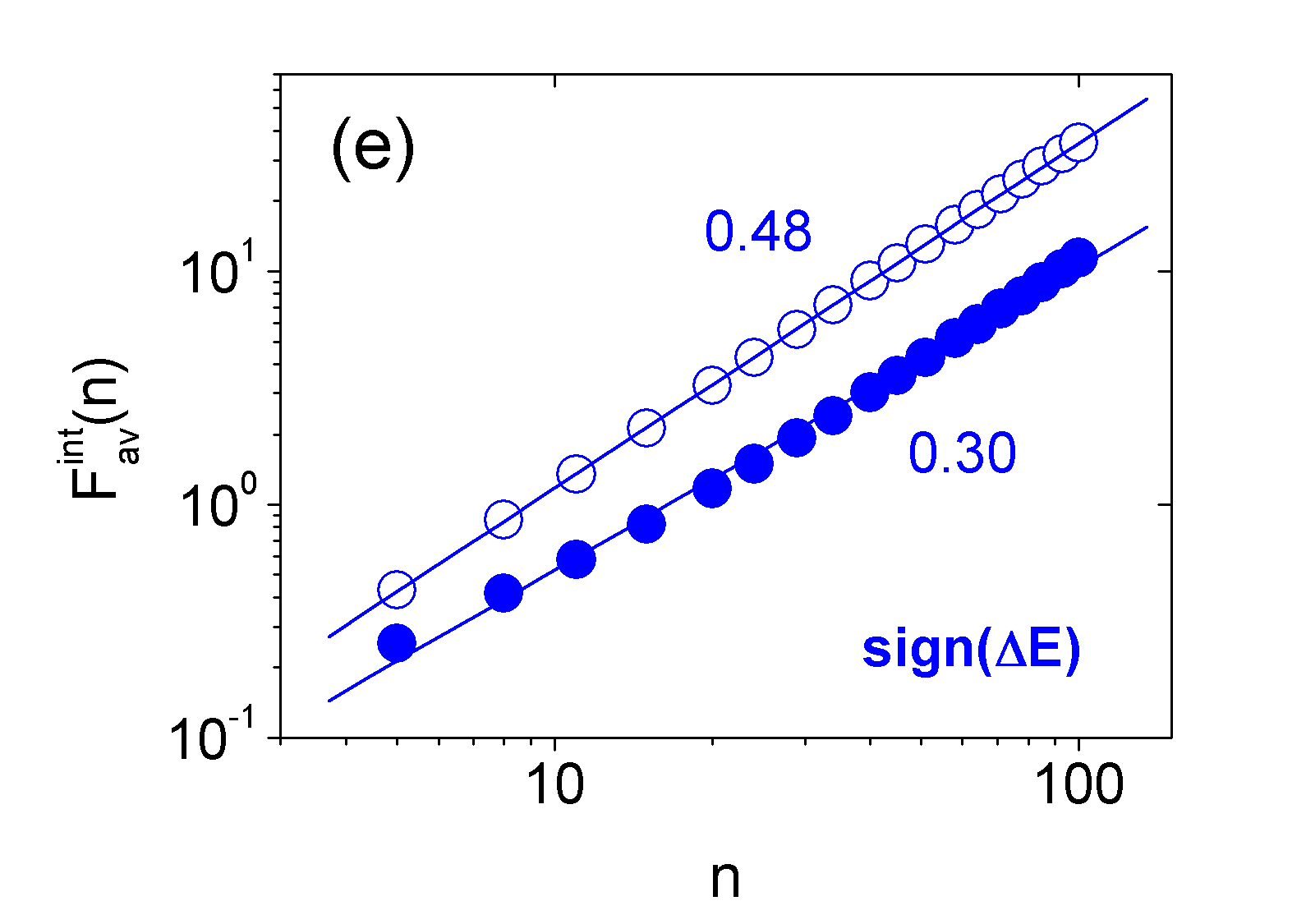}
\includegraphics[scale=.22]{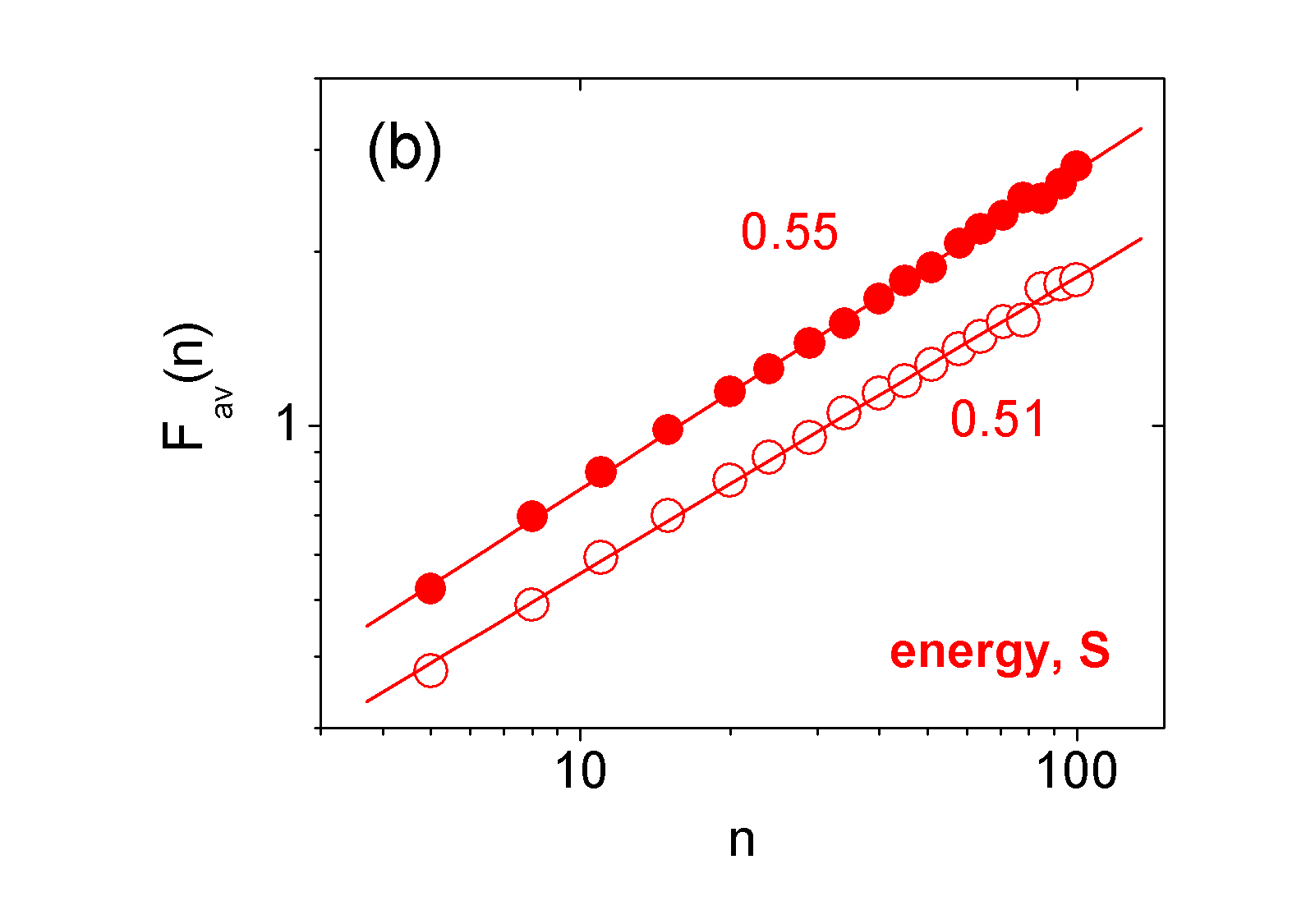}
\includegraphics[scale=.22]{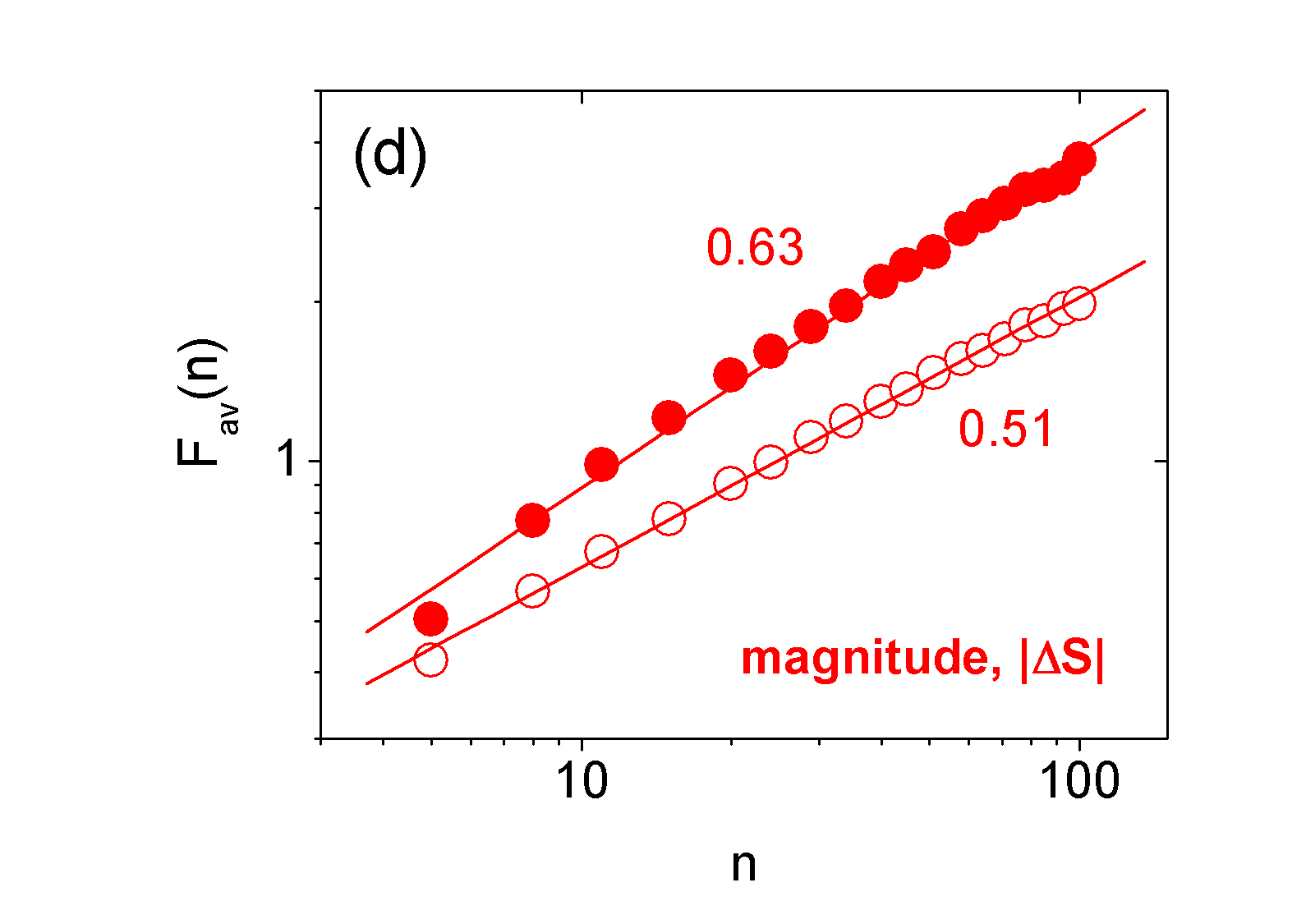}
\includegraphics[scale=.22]{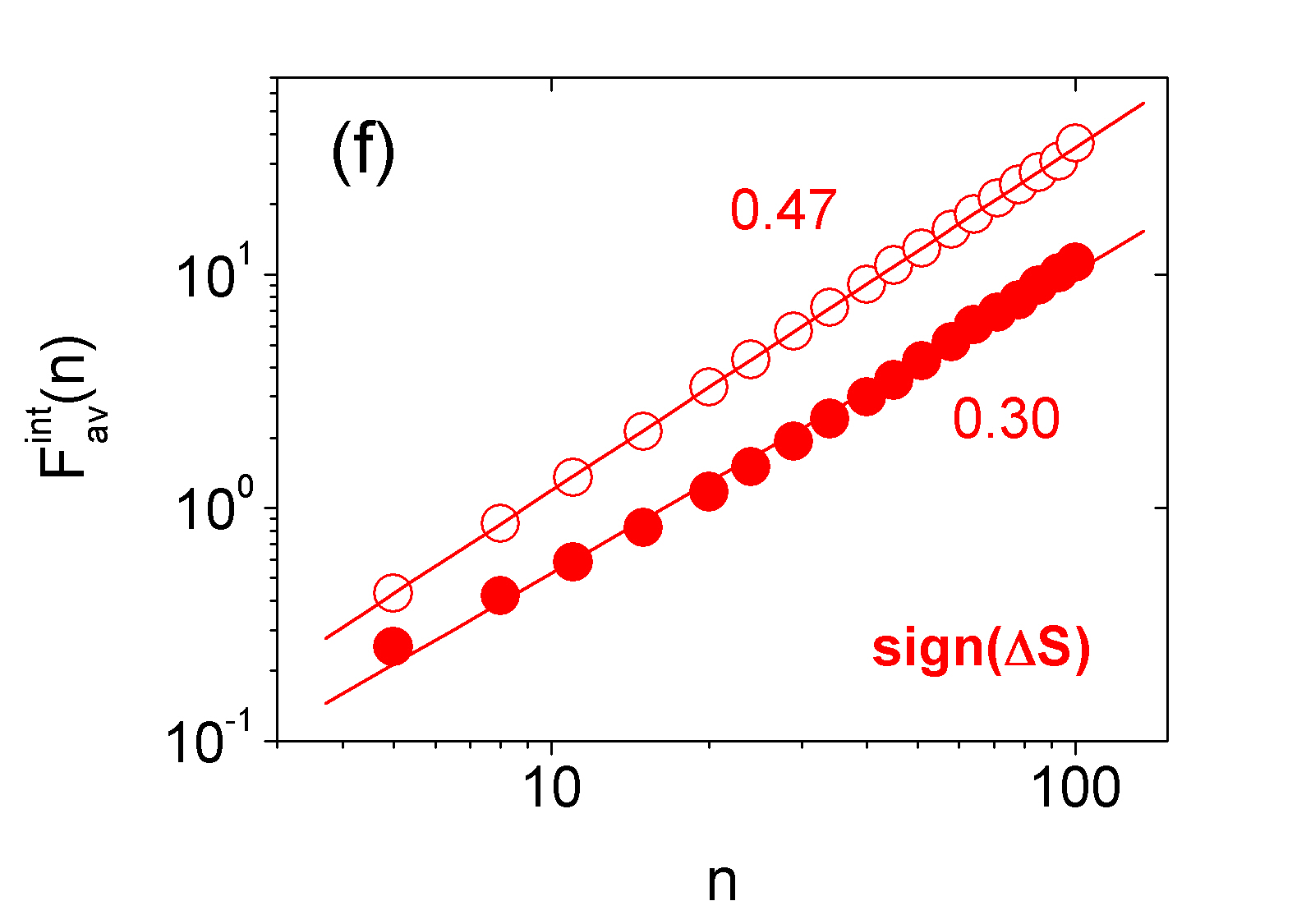}
 \caption{Analysis of temporal correlations in the data. $F_{av}(n)$ is an
 average of the DFA fluctuation function $F(n)$ calculated for several subseries.
$F_{av}^{int}(n)$ is the average DFA fluctuation function
 obtained from integrated subseries. We investigate $F$ in the range $5\leq n \leq
 100$. We also show $F_{av}$ and $F_{av}^{int}$
obtained from shuffled series (open circles). In all cases, the
exponent $h$ is obtained by least square linear fits to the data. As
expected, $h \simeq 0.5$ for all shuffled series. {\bf (a)}
$F_{av}(n)$ calculated from normalized energy series for
earthquakes, $E/\sigma(E)$. The slope gives $h=0.57$. {\bf (b)}
$F_{av}(n)$ calculated from normalized energy series for crumpled
plastic sheets, $S/\sigma(S)$. The linear fits gives $h=0.55$. {\bf
(c)} $F_{av}(n)$ calculated from normalized magnitude series for
earthquakes, $|\Delta E|/\sigma(|\Delta E|)$, giving $h=0.66$. {\bf
(d)} $F_{av}(n)$ calculated from normalized magnitude series for
crumpled plastic sheets, $|\Delta S|/\sigma(|\Delta S|)$. In this
case, $h=0.63$. {\bf (e)} $F_{av}^{int}(n)$ calculated from
integrated sign series for earthquakes, $\mbox{sign}(\Delta E)$. The
slope is $1.30$, giving $h=0.30$. {\bf (f)} $F_{av}^{int}(n)$
calculated from integrated sign series for crumpled plastic sheets,
$\mbox{sign}(\Delta S)$. The slope is also $1.30$, giving $h=0.30$.}
\label{fig4}
\end{figure*}

Next, we investigate fractal properties in the output signal of the
systems. For a given time series $x(i)$, the autocorrelation
function is defined as $C(\tau)=\langle x(i)x(i+\tau)
\rangle-\langle x(i)^{2} \rangle$. In order to reduce fluctuations
in $C(\tau)$ it is common to obtain the root mean square fluctuation
function $F(\tau)$, such that
$F(\tau)^{2}=\sum_{i=1}^{\tau}\sum_{j=1}^{\tau}C(j-1)$\cite{r12}.
The net displacement after $\tau$ steps is
$y(\tau)=\sum_{i=1}^{\tau}x(i)$ and the root mean square fluctuation
is defined as $\widehat{F}(\tau)=\sqrt{\langle \Delta y(\tau)^{2}
\rangle-\langle \Delta y(\tau) \rangle^{2}}$, where $\Delta
y(\tau)=y(\tau_{0}+\tau)-y(\tau_{0})$. For fractal series,
$\widehat{F}(\tau)$ follows a power law behavior,
$\widehat{F}(\tau)\sim \tau^{h}$, where $h$ is the scaling exponent
which quantifies the degree of correlations. When $h>0.5$ ($h<0.5$)
the series is long-range correlated (anti-correlated). Uncorrelated
series present $h=0.5$. Short-range correlations also may exhibit
$h=0.5$.

For nonstationary records, it is common to apply detrended
fluctuation analysis (DFA)\cite{peng94,buldyrev95} to investigate
correlations in the data. For fractal series, the DFA root mean
square fluctuation, $F(n)$, also follows a power law behavior,
\begin{equation}\label{eq4}
    F(n)\sim n^{h},
\end{equation}
where $n$ is a time scale. Here we apply DFA method in order to
quantify temporal correlations in the data.

The typical size of a given time series in the experiment of
crumpled plastic sheets is $\sim 1016$ (20 samples). In order to
perform a parallel analysis, we partition the energy series for
earthquakes in subseries of size $1016$ (428 samples). For each
subseries we obtain the DFA root mean square fluctuation, $F(n)$,
and perform an average of $F(n)$ over all subseries - obtaining
$F_{av}(n)$. Because the size of a typical subseries, we investigate
$F(n)$ in the range $5<n<100$. Figures 3a and 3b shows $F_{av}(n)$
for the energy series for earthquakes and crumpled plastic sheets -
$E$ and $S$. We find $h\simeq 0.55$ for both records suggesting weak
correlations in the data. As expected, we find $h\simeq 0.5$ for
shuffled series.

Starting from the sequence of energy increments, we also obtain two
sub-series: magnitude of energy increments - $|\Delta E|$ and
$|\Delta S|$ - and sign of energy increments - $\mbox{sign}[\Delta
E]$ and $\mbox{sign}[\Delta S]$. The function $\mbox{sign}[\Delta
x]$ assumes the values $-1$, $0$ or $1$ if the increment $\Delta x$
is negative, null or positive, respectively. For details of the
magnitude-sign decomposition approach, see refs.
\cite{ashkenazy01,ashkenazy03}. Figures 3c and 3d shows $F_{av}(n)$
for the magnitude series of energy increments for earthquakes and
plastic sheets. For both records $h\simeq 0.65$ indicating long
range correlations in the data. Figures 3e and 3f shows
$F_{av}^{int}(n)$ - the average DFA fluctuation function obtained
from integrated series - for the sign series of energy increments
for earthquakes and plastic sheets. This previous integration is
necessary in DFA method when $h<0.5$. For both records we find
$h\simeq 0.30$ indicating anti-correlations in the data. As
expected, shuffled magnitude and sign series exhibits $h \simeq 0.5$
indicating uncorrelated behavior. Notice the quantitative agreement
between the values of $h$ for both systems.

The analysis reported here indicates remarkable similarities between
two distinct out-of-equilibrium physical systems, providing a
quantitative support for the analogy between earthquakes and
crumpled films. Specifically, we show that for both signals i) the
distribution of energy follows a power law with exponent $\alpha
\simeq 3$; ii) the distribution of energy increments exhibits a
common non-Gaussian shape in the range $1 \leq \tau \leq 100$, with
power law tails with exponent $\beta \simeq 3$; iii) the
distribution of return intervals follows an exponential behavior;
iv) the DFA power law exponent is $h\simeq 0.55$ for energy series,
$h\simeq 0.65$ for magnitude series of energy increments and
$h\simeq 0.30$ for sign series of energy increments in the range $5
\leq n \leq 100$. These findings are consistent with the hypothesis
that earthquakes and crumpled plastic sheets may be driven by common
underlying mechanisms.

The nature of both processes analyzed here also presents analogies.
It has been pointed that the energy stored in a crumpled material is
originated in the buckling process while the film is crumpled, and
it is mainly concentrated in the formed ridges\cite{r14,r15}. The
nonequilibrium behavior observed in the relaxation process can be
understood as a consequence of the frustration in the crossed
ridges, characterizing a stress\cite{r16}. Moreover, it is known
that Earth´s crust can also exhibit buckling under viscous stresses
on its layers\cite{r15}.

Some phenomenological models, as the epidemic-type aftershock
sequence model (ETAS)\cite{r17,r18}, the continuous-time random walk
models (CTRW)\cite{r18} and the Olami-Feder-Christensen model
(OFC)\cite{caruso07,r19,r20} try to incorporate the main properties
of the complex spatiotemporal behavior of earthquakes. Since the
experiments with crumpled plastic sheets are simple and
reproducible, they may be used as an additional data source to
compare with artificial data. We hope that this analogy could
provide some insight on the mechanisms behind the complex spatial
and temporal behavior of earthquakes.

\acknowledgments

We thank CNPq (Brasilian Agency) for partial financial support. We
also thank P. Palffy-Muhoray for discussions.

\end{document}